\documentstyle[12pt]{article}
\newcommand\be{\begin{equation}}
\newcommand\ee{\end{equation}}
\newcommand\bea{\begin{eqnarray}}
\newcommand\eea{\end{eqnarray}}
\newcommand\Pp{{\cal{P}}}
\newcommand\G{{\cal{G}}}

\newcommand\Pm{\widetilde{\cal{P}}}
\newcommand\const{{\rm{const}}}
\newcommand\Sp{{\rm{Sp\,}}}

\newcommand\e{{\rm{e}}}
\newcommand\rv{{\bf r}}
\newcommand\op{{\!\!\!\!\!\!}}
\newcommand\Q{{\cal{Q}}}
\newcommand\Pd{{\textstyle{1\over2}}}

    \def\newpic#1{%
    \def\emline##1##2##3##4##5##6{%
       \put(##1,##2){\special{em:point #1##3}}%
       \put(##4,##5){\special{em:point #1##6}}%
       \special{em:line #1##3,#1##6}}}
 \newpic{}

\begin{document}
\
\vspace{0.5cm}

\begin{center}
{\Large\bf THE CORRELATION FUNCTION\\
IN TWO DIMENSIONAL
\vspace{.2cm}
ISING MODEL\\ ON THE FINITE SIZE LATTICE. I.}

\bigskip
 A.I.~Bugrij
\footnote
{e-mail:  abugrij@bitp.kiev.ua },

\medskip
 {\it Bogolyubov Institute for Theoretical Physics}\\
 \medskip
 {\it 03143 Kiev-143, Ukraine}
\end{center}

\bigskip
\begin{abstract}
\begin{sloppypar}
 The correlation function of two dimensional Ising model with the
 nearest neighbours interaction on the finite size lattice with the
periodical boundary conditions is derived. The expressions similar to
 the form factor representation are obtained both for the
 ferromagnetic and paramagnetic regions of coupling constant.
The peculiarities caused by the finite size of the system are
discussed.  The asymptotic behaviour of the corresponding quantities
as functions of size is calculated. Some generalization of the
obtained result is conjectured.

\vspace{1cm}
 \noindent
  PACS number(s): 05.50.+q, 11.10.-z
\end{sloppypar} \end{abstract} \thispagestyle{empty}

\newpage
\renewcommand{\theequation}{1.\arabic{equation}}
\setcounter{equation}{0}

\section{Introduction}

Since the outstanding result of Montroll, Potts, Ward [1] there
appeared many papers devoted to the problem of the spin-spin
correlation function in the two dimensional Ising model (IM) with
the nearest-neighbour interaction on the infinite size lattice
(see e.g. refs. in [2]). The basic IM theory results are collected
in the significant book [3]. I can not, of course, to give here
any comprehensive list of IM references. Only very few of them
which were straight instructive for the preparation of the present
paper will be referred to. I think, in particular, that the
results and ideas of the paper [4] were very fruitful for the
further development of the IM theory. The study of the scaling
limit of the IM (see e.g. [4]--[7]) is of interest from the
quantum field theory point of view.

The so-called form factor representation [7] for the two
dimensional IM correlation function  appears to be a crown of this
activity:  \be
\langle\sigma(\rv_1)\sigma(\rv_2)\rangle=\const\sum_{n}g_n(r),\quad
r=|\rv_1-\rv_2|, \ee
\be
g_n(r)={1\over
n!(2\pi)^n}\int\limits_{-\infty}^{\infty}\prod_{i=1}^{n}
\biggl({dq_i\over\omega_i}\e^{-r\omega_i}\biggr)F_n^2[q], \ee \be
F_n[q]=\prod_{i<j}^{n}\biggl({q_i-q_j\over\omega_i+\omega_j}\biggr),
\quad \omega_i=\omega(q_i)=\sqrt{m^2+q_i^2}. \ee In the
ferromagnetic region of the coupling parameter the summation in
(1.1) is extended over even $n$, in the paramagnetic  --- over
odd. It is worth of noting that this representation  was first
evaluated [7] in the framework of the $S$-matrix approach [8] and
then [6] by means of straightforward IM solution. The discovery of
the form factor representation was very fruitful for the progress
of exactly integrable quantum field theories
 [9].

The advantage of representation (1.1) consists in that the
dynamical aspects of the system are separated from the kinematical
ones: form factors squared are integrated over the phase volumes of
the $n$-particle intermediate states. The form factor representation
 (1.1) clarifies the dynamics of the model but does not answer on
the questions about the spectrum, sort and statistics of the
particles that form the intermediate and asymptotic states of the
system under consideration. The analysis of the model at finite
(nonzero) temperature or in the finite volume is necessary for
this purpose.  One can observe an activity in this field in last
years [10]--[16].  These works show in particular that the problem
is complicated enough:  the authors of refs.  [15], [16],  for
example, call the conjectures of [12], [14] in question. I think
that a simple exactly solvable lattice model example would be very
useful for the business.  Formally the finite temperature (in
quantum field theory sense) means the finite size along the
temporal axis and periodical boundary condition for boson fields
or antiperiodical --- for fermion fields. Meanwhile there is no
representation which is analogous to the form factor one for the
correlation function on the finite size lattice even for the Ising
model.

This work, I hope, makes up a deficiency. I have calculated the IM
spin-spin correlator on the lattice wrapped on a cylinder by the use
of the classic methods of IM theory [3] adapted properly to the case
of the finite sizes. The solution is expressed in the form similar to
the form factor expansion (1.1)--(1.3). For reader's convenience I
write the result just in the Introduction. If one considers it obvious
he would be free of cumbersome mathematical
transformations. So, the
expression obtained in this work is  \be
\langle\sigma(\rv_1)\sigma(\rv_2)\rangle=\xi\xi_T\e^{-r/\Lambda}
\sum_{n}g_n(r), \ee \be g_n(r)={\e^{-n/\Lambda}\over
n!(N)^n}{\sum_{[q]}}^{(b)} \prod_{i=1}^{n}
\biggl({\e^{-r\gamma_i-\eta_i}\over\sinh \gamma_i}
\biggr)F_n^2[q], \ee \be
F_n[q]=\prod_{i<j}^{n}{\sin((q_i-q_j)/2)\over\sinh((\gamma_i+\gamma_j)/2)}.
\ee
More precisely, the representation (1.4)--(1.6) corresponds to the
correlation of spins posed on the line parallel to the axis of a
        cylinder with $N$ sites on its circumference.  The values $\xi$,
$\xi_T$, $\Lambda$, $\gamma_i$, $\eta_i$, in eqs.  (1.4)--(1.6) are
 defined further in Sections 2, 3, 4.  Three of them --- $\xi_T$,
 $\Lambda$ and $\eta_i$ --- are specific cylinder circumference
dependent values:  $\ln\xi_T$, $\Lambda^{-1}$, $\eta_i$ vanish if
$N$ tends to infinity. The appearance of the summation in (1.5)
instead of the integration over phase volume is natural
consequence of the size finiteness. It is widely believed that the
underlying IM field is the fermion one: for instance the IM
partition function is exactly the same as in the free Majorana
fermion system. So, the boson spectrum of quasimomentum in
Brillouin zone (this is denoted by the upper index in sum (1.5))
is surprising . Nevertheless it follows unambiguously         from
the calculations. It takes attention also
 that the lattice form factor (1.6) does not depends explicitly on the
 cylinder circumference. Note that similar expression for  $F_n[q]$ at
even $n$ was found by authors of  [6] for the infinite lattice
case.

In Section 2 the model is formulated and the brief evaluation of
the Toeplitz determinant representation for the IM correlation
function is given. The corresponding Toeplitz matrix allowing for
the finite size lattice is calculated. In Section 3 the lattice
form factor representation for the correlation function is derived
for the ferromagnetic domain of the coupling parameter and for the
paramagnetic domain --- in Section 4.
In Conclusion I discuss the relationship
between the IM correlation function on a cylinder and quantum
field Green function at finite temperature or in finite volume. I
also comment on some prospect of generalization of the obtained
results. In Appendix the factorised representations for the
Toeplitz determinants of special form which are used for obtaining
lattice form factor expansions are evaluated.

\renewcommand{\theequation}{2.\arabic{equation}}
\setcounter{equation}{0}
\section{The Model}

The Ising model with the nearest neighbour interaction on the square
 $M\times N$ lattice (see. Fig. 1) is defined by the hamiltonian
 $H[\sigma]$
$$
-\beta
H[\sigma]=K\sum_{\bf{r}}\sigma({\bf
r})(\nabla_x+\nabla_y)\sigma({\bf r}), \quad K=\beta J, $$
where $J$ is the coupling parameter, $\beta$ is the inverse
temperature; two dimensional vector ${\bf r}=(x,y)$ numerates the
lattice sites:  $x=1,\,2,\,\ldots,\,M$,\ \ $y=1,\,2,\,\ldots,\,N$;\ \
$\nabla_x$, $\nabla_y$ are the one step shift operators
$$
\nabla_x\sigma(x,y)=\sigma(x+1,y),\quad
\nabla_y\sigma(x,y)=\sigma(x,y+1).
$$

\begin{center}
\newpic{1}
\special{em:linewidth 0.4pt}
\unitlength 1.00mm
\linethickness{1.0pt}
\begin{picture}(113.00,80.00)
\emline{10.00}{10.00}{1}{10.00}{70.00}{2}
\emline{10.00}{70.00}{3}{100.00}{70.00}{4}
\emline{100.00}{70.00}{5}{100.00}{10.00}{6}
\emline{100.00}{10.00}{7}{10.00}{10.00}{8}
\emline{20.00}{10.00}{9}{20.00}{70.00}{10}
\emline{40.00}{10.00}{11}{40.00}{70.00}{12}
\emline{50.00}{10.00}{13}{50.00}{70.00}{14}
\emline{60.00}{10.00}{15}{60.00}{70.00}{16}
\emline{70.00}{10.00}{17}{70.00}{70.00}{18}
\emline{90.00}{10.00}{19}{90.00}{70.00}{20}
\emline{100.00}{20.00}{21}{10.00}{20.00}{22}
\emline{100.00}{30.00}{23}{10.00}{30.00}{24}
\emline{100.00}{50.00}{25}{10.00}{50.00}{26}
\emline{100.00}{60.00}{27}{10.00}{60.00}{28}
\put(30.00,40.00){\circle{2.00}}
\put(80.00,40.00){\circle{2.00}}
\put(0.00,80.00){\vector(0,1){0.2}}
\emline{0.00}{0.00}{29}{0.00}{80.00}{30}
\put(110.00,0.00){\vector(1,0){0.2}}
\emline{0.00}{0.00}{31}{110.00}{0.00}{32}
\emline{0.00}{70.00}{33}{1.00}{70.00}{34}
\emline{0.00}{60.00}{35}{1.00}{60.00}{36}
\put(-5.00,70.00){\makebox(0,0)[cc]{$N$}}
\put(25.00,45.00){\makebox(0,0)[cc]{$\sigma(\rv_1)$}}
\put(85.00,45.00){\makebox(0,0)[cc]{$\sigma(\rv_2)$}}
\emline{0.00}{50.00}{37}{1.00}{50.00}{38}
\emline{0.00}{40.00}{39}{1.00}{40.00}{40}
\emline{0.00}{30.00}{41}{1.00}{30.00}{42}
\emline{0.00}{20.00}{43}{1.00}{20.00}{44}
\emline{0.00}{10.00}{45}{1.00}{10.00}{46}
\emline{10.00}{0.00}{47}{10.00}{1.00}{48}
\emline{10.00}{1.00}{49}{10.00}{1.00}{50}
\emline{10.00}{1.00}{51}{10.00}{2.00}{52}
\emline{20.00}{0.00}{53}{20.00}{1.00}{54}
\emline{20.00}{1.00}{55}{20.00}{1.00}{56}
\emline{20.00}{1.00}{57}{20.00}{2.00}{58}
\emline{30.00}{0.00}{59}{30.00}{1.00}{60}
\emline{30.00}{1.00}{61}{30.00}{1.00}{62}
\emline{30.00}{1.00}{63}{30.00}{2.00}{64}
\emline{40.00}{0.00}{65}{40.00}{1.00}{66}
\emline{40.00}{1.00}{67}{40.00}{1.00}{68}
\emline{40.00}{1.00}{69}{40.00}{2.00}{70}
\emline{50.00}{0.00}{71}{50.00}{1.00}{72}
\emline{50.00}{1.00}{73}{50.00}{1.00}{74}
\emline{50.00}{1.00}{75}{50.00}{2.00}{76}
\emline{60.00}{0.00}{77}{60.00}{1.00}{78}
\emline{60.00}{1.00}{79}{60.00}{1.00}{80}
\emline{60.00}{1.00}{81}{60.00}{2.00}{82}
\emline{70.00}{0.00}{83}{70.00}{1.00}{84}
\emline{70.00}{1.00}{85}{70.00}{1.00}{86}
\emline{70.00}{1.00}{87}{70.00}{2.00}{88}
\emline{80.00}{0.00}{89}{80.00}{1.00}{90}
\emline{80.00}{1.00}{91}{80.00}{1.00}{92}
\emline{80.00}{1.00}{93}{80.00}{2.00}{94}
\emline{90.00}{0.00}{95}{90.00}{1.00}{96}
\emline{90.00}{1.00}{97}{90.00}{1.00}{98}
\emline{90.00}{1.00}{99}{90.00}{2.00}{100}
\emline{100.00}{0.00}{101}{100.00}{1.00}{102}
\emline{100.00}{1.00}{103}{100.00}{1.00}{104}
\emline{100.00}{1.00}{105}{100.00}{2.00}{106}
\put(-5.00,-5.00){\makebox(0,0)[cc]{0}}
\put(-5.00,40.00){\makebox(0,0)[cc]{$y$}}
\put(-5.00,10.00){\makebox(0,0)[cc]{1}}
\put(30.00,-5.00){\makebox(0,0)[cc]{$x$}}
\put(10.00,-5.00){\makebox(0,0)[cc]{1}}
\put(20.00,-5.00){\makebox(0,0)[cc]{$\ldots$}}
\put(-5.00,25.00){\makebox(0,0)[cc]{$\vdots$}}
\put(80.00,-5.00){\makebox(0,0)[cc]{$x+r$}}
\put(100.00,-5.00){\makebox(0,0)[cc]{$M$}}
\put(90.00,-5.00){\makebox(0,0)[cc]{$\ldots$}}
\put(55.00,-5.00){\makebox(0,0)[cc]{$\ldots$}}
\put(-5.00,55.00){\makebox(0,0)[cc]{$\vdots$}}
\put(50.00,-15.00){\makebox(0,0)[cc]
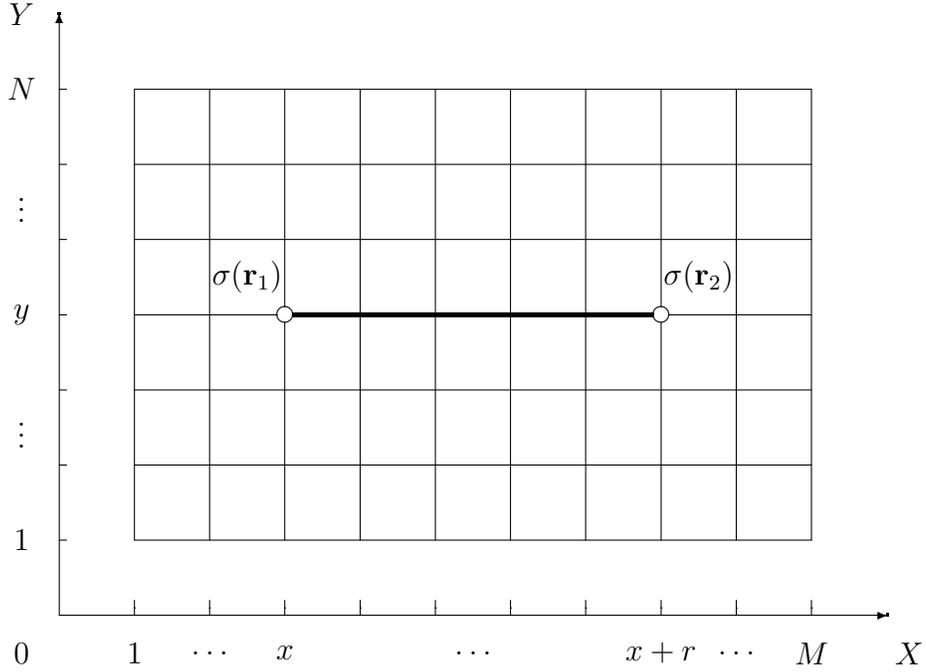
{\ \hspace{2cm} {Fig. 1. {\small The positions of the correlating
Ising spins on the lattice.}}}}
\put(-5.00,80.00){\makebox(0,0)[cc]{$Y$}}
\put(113.00,-5.00){\makebox(0,0)[cc]{$X$}}
\emline{30.00}{41.00}{107}{30.00}{70.00}{108}
\emline{30.00}{39.00}{109}{30.00}{10.00}{110}
\emline{80.00}{39.00}{111}{80.00}{10.00}{112}
\emline{80.00}{41.00}{113}{80.00}{70.00}{114}
\emline{29.00}{40.00}{115}{10.00}{40.00}{116}
\emline{10.00}{40.00}{117}{10.00}{40.00}{118}
\emline{81.00}{40.00}{119}{100.00}{40.00}{120}
\linethickness{1.4pt}
\put(31.00,40.00){\line(1,0){48.00}
\linethickness{0.4pt}}
\end{picture}
\end{center}
\vspace{2cm}

\noindent
The partition function of the model is
\be
Z=\sum_{[\sigma]}\e^{-\beta H[\sigma]},
\ee
the pair correlation function is
\be
\langle\sigma({\bf r}_1)\sigma({\bf r}_2)\rangle=Z^{-1}
\sum_{[\sigma]}\e^{-\beta H[\sigma]}\sigma({\bf r}_1)\sigma({\bf
r}_2). \ee

For the lattice with the periodical boundary conditions (wrapped on a
torus) the partition function (2.1) can be expressed [17], [18]
through the combination of the four Gauss Grassmann functional
integrals with different boundary conditions \be
Z=(2\cosh^2
K)^{MN}{\textstyle{1\over2}}(\Q^{(ff)}+\Q^{(bf)}+\Q^{(fb)}-\Q^{(bb)}),
\ee
where
$$
\Q=\int d[\psi]\e^{S[\psi]},\qquad
d[\psi]=\prod_{\bf r}
d\psi^1({\bf r})
d\psi^2({\bf r})
d\psi^3({\bf r})
d\psi^4({\bf r}).
$$
The action $S[\psi]$ is the anticommuting quadratic form
\be
S[\psi]=\Pd\sum_{{\bf r},{\bf
r}'}\sum_{\nu,\rho=1}^{4}\psi^\nu({\bf r})
D^{\nu\rho}({\bf
r},{\bf r}')\psi^\rho({\bf r}')=\Pd(\psi\hat{D}\psi), \ee
\be
\hat{D}=
\left(\begin{array}{cccc}
0&1+t\nabla_x&1&1\\
-1-t\nabla_{-x}&0&-1&1\\
-1&1&0&1+t\nabla_{y}\\ -1&-1&-1-t\nabla_{-y}&0
\end{array}\right),
\quad t=\tanh K.  \ee
The indices $(f,b)$ in eq. (2.3) shows the type of the boundary
conditions for the shift operators in the matrix (2.5).
\begin{itemize}
\item[``$f$" --]
antiperiodical:  $(\nabla_x^{(f)})^M =-1$,
$(\nabla_y^{(f)})^N =-1$;\\ the corresponding quasimomentum runs over
halfinteger values in Bril\-lou\-in zone\\ $(-\pi,\pi)$
(in ${2\pi/M}$ unites for
$p_x$ and ${2\pi/N}$ -- for  $p_y$)
\item[``$b$" --] periodical:
$(\nabla_x^{(b)})^M =1$, $(\nabla_y^{(b)})^N =1$;\\ the corresponding
components of quasimomentum runs over integer values.
\end{itemize}

The correlation function (2.2) can be also presented in terms of the
Grassmann functional integrals with the Gauss distribution [18]
\be
\langle\sigma({\bf r}_1)\sigma({\bf r}_2)\rangle=
t^r{\Q_d^{(ff)}+\Q_d^{(bf)}+\Q_d^{(fb)}-\Q_d^{(bb)}\over\Q^{(ff)}+
\Q^{(bf)}+  \Q^{(fb)}-  \Q^{(bb)}},
\ee
where  $\Q_d$ is the functional integral
\be
\Q_d=\int d[\psi]\e^{S_d[\psi]}.
\ee
The action with the defect  denoted by  $S_d[\psi]$ in (2.7) differs
from (2.4) in that the parameter  $t$  is replaced
by $t^{-1}$ on the links along the path which connects the sites ${\bf
 r}_1$ and  ${\bf r}_2$.

I consider the special case when the sites
${\bf r}_1$ and  ${\bf r}_2$ are situated on the line parallel to the
horizontal axis as is shown on Fig. 1:
$$
{\bf r}_1=(x,y),\quad {\bf r}_2=(x+r,y),\quad |\rv_1-\rv_2|=r .
$$
The action with the defect in this case has the form
$$ S_d[\psi]=S[\psi]+\Delta S[\psi],$$
where
$$ \Delta S[\psi]=(t^{-1}-t)
\sum_{x'=1}^{r}\psi^1(x+x'-1,y)\psi^2(x+x',y).$$
The ratio in the r.h.s. of (2.6) simplifies
\be
\langle\sigma({\bf r}_1)\sigma({\bf r}_2)\rangle=
t^r{\Q_d^{(ff)}\over \Q^{(ff)}}[1+o(e^{-M/N})],
\ee
if at least one of the lattice sizes is much larger than other, i.e.
torus degenerates to cylinder: in our case  $M/N\gg1$. Note,
 that only the terms with the antiperiodical boundary
conditions survive.

After some manipulations with the functional integrals the ratio in
the r.h.s. of (2.8) can be transformed to the functional integral over
Grassmann field which is defined just on the line connecting the sites
${\bf
r}_1$ and ${\bf r}_2$
\bea &&\langle\sigma({\bf r}_1)\sigma({\bf
r}_2)\rangle= \int d[\chi]\e^{(\chi^1A\chi^2)}=\det A,\\
&&d[\chi]=\prod_{x=0}^{r-1}d\chi^1_x\,d\chi^2_x,\quad
(\chi^1A\chi^2)=\sum_{x,x'}\chi^1_xA_{x,x'}\chi^2_{x'},
\nonumber\\
&&A_{x,x'}={1\over MN}{\sum_{{\bf p
}}}^{(f)}\ {\e^{ip_x(x-x')}
[2t(1+t^2)-(1-t^2)(\e^{ip_x}+t^2\e^{-ip_x})]\over(1+t^2)^2-2t(1-t^2)
(\cos p_x+\cos p_y)},\\
&&\ \qquad x,x'=0,\,1,\,\ldots,\,r-1.
\nonumber
\eea
I recall that the index $(f)$ points the antiperiodicity: the
momentum's components run over halfinteger values  --
fermion spectrum  $$
{\sum_{p_x}}^{(f)}v(p_x)=\sum_{l=1}^{M}v\biggl({2\pi\over
M}(l+\Pd)\biggr),\quad
{\sum_{p_y}}^{(f)}v(p_y)=\sum_{l=1}^{N}v\biggl({2\pi\over
N}(l+\Pd)\biggr).  $$

Now let us rewrite the expression  (2.10) in appropriate for
the Toeplitz determinant theory manner. From the tabulated
formulas [20]
\be {\prod_{p_y}}^{(f)}2(\cosh\gamma-\cos
p_y)=\e^{N\gamma}(1+\e^{-N\gamma})^2, \quad {\prod_{p_y}}^{(b)}2
(\cosh\gamma-\cos p_y) =\e^{N\gamma}(1-\e^{-N\gamma})^2, \ee
it follows in particular \be
{1\over N}{\sum_{p_y}}^{(f)}{1\over \cosh\gamma-\cos p_y}=
{\tanh(N\gamma/2)\over\sinh\gamma},
\quad{1\over N}{\sum_{p_y}}^{(b)}{1\over \cosh\gamma-\cos p_y}=
{\coth(N\gamma/2)\over\sinh\gamma}.
\ee
At the cylinder case $M\to\infty$, $N=\const$ the summation over
$p_x$ turns into the integral
 \be {1\over
M}{\sum_{p_x}}^{(f)}v(\e^{-ip_x})={1\over2\pi
i}\oint\limits_{|z|=1}{dz\over z} v(z)+o(\e^{-M\mu}).
\ee
Summing up
 (2.10)
over
$p_y$ by use of
(2.12) and accounting for (2.13) one can obtain for the matrix
   $A_{x,x'}$
   \be A_{x,x'}={1\over2\pi i}\oint\limits_{|z|=1}
{dz\over z}z^{-x+x'}A(z), \ee where the kernel $A(z)$ is \be
A(z)=\sqrt{{(1-\e^{-\gamma_\pi}z)(1-\e^{-\gamma_0}z^{-1})\over
(1-\e^{-\gamma_0}z)(1-\e^{-\gamma_\pi}z^{-1})}}\cdot T(z),\quad
z=\e^{-ip},
\ee
and
\be
T(z)=\tanh[(N\gamma(p))/2],
\ee
\be
\gamma_0=|2K+\ln t|,\quad \gamma_\pi=2K-\ln t.
\ee
The function $\gamma(p)$ in  (2.16) is defined by the following
equation
\be 4\sinh^2{\gamma(p)\over2}=\mu^2+4\sin^2{p\over2},
\ee
where
\be
\mu^2 =2(\sqrt{\sinh 2K}-1/\sqrt{\sinh 2K})^2.  \ee
It follows from eqs. (2.18), (2.19) in particular $$
\gamma(0)=\gamma_0,\quad\gamma(\pi)=\gamma_\pi,\quad
\cosh\gamma_\pi-\cosh\gamma_0=2.$$

The function $T(z)$ (2.16) tends to unity at the thermodynamic limit
$N\to\infty$ (more precise $N\mu\gg1$), and
  the kernel $A(z)$ (2.15) turns into the classic IM theory
expression. In this limit the value $K=K_c$ is the critical point:
the specific heat and correlation length diverge. At the finite $N$
the corresponding singularities are smoothed and there is no phase
transition. Nevertheless also in this case the value $K=K_c$ is
notable.
 The calculation technique based on
Wiener--Hopf sum equation  ``distinguishes'' the ferro- and
paramagnetic regions and
the expression for the correlation function at $K>K_c$
differs from that at
 $K>K_c$.
 Just in this sense I shall mean further the
words ``critical point'', ``phase'' and so on. Note that the following
equalities take place at the critical point
$$
\sinh 2K_c=1,\quad \mu=0, \quad \gamma_0=0,\quad
\gamma_\pi=2\ln(\sqrt{2}+1).
$$

\renewcommand{\theequation}{3.\arabic{equation}}
\setcounter{equation}{0}

\section
{The Ferromagnetic region: ${\bf K>K_c}$}

The dependence of the correlation function (2.9)
\be
\langle\sigma(\rv_1)\sigma(\rv_2)\rangle=\det A^{(r)}
\ee
on the distance
$r=|\rv_1-\rv_2|$
enters through the matrix dimension denoted by upper index $(r)$
in the r.h.s. of (3.1). The dependence on the cylinder circumference
 $N$ comes into the function $T(z)$ (2.16) and the dependence on the
coupling parameter $K$ is defined by the parameter $\mu$ (2.19) in the
eq.  (2.18) for
 the function
 $\gamma(p)$.

The kernel (2.14) of the matrix (2.9) $A^{(r)}$
has to be represented in the factorized form according to the method
of the determinant calculation which we use  (see (A6), (A7) in the
Appendix),
$$ A(z)=P(z)Q(z^{-1}),
$$
where the functions  $P(z)$ and  $Q(z)$ are analytic ones inside the
$|z|=1$ circle.  It can be formally done by use of the projection
operators (A8) that are defined in the Appendix \bea 2\ln
P(z)&=&\ln(1-\e^{-\gamma_\pi}z)-\ln(1-\e^{-\gamma_0}z)+2\Pp \ln T(z),
\nonumber\\ 2\ln
Q(z^{-1})&=&\ln(1-\e^{-\gamma_0}z^{-1})-\ln(1-\e^{-\gamma_\pi}z^{-1})+2\Pm
\ln T(z).
\nonumber
\eea
Otherwise accounting for the explicit form (2.16) one can express
$T(z)$ through the ratio of the products with help of eqs. (2.11)
\be
T^2(z)={{\prod\limits_{q}}^{(b)}[\cosh\gamma(p)-\cos
q]\over{\prod\limits_{q}}^{(f)}[\cosh\gamma(p)-\cos q]}.  \ee It
follows from (2.18) $$ \cosh\gamma(p) -\cos
q=2+{\mu^2\over2}-\cos q-\cos
p=\Pd\e^{\gamma(q)}(1-\e^{-\gamma(q)}z)(1-\e^{-\gamma(q)}z^{-1}), $$
so eq. (3.2) turns into the form \be T^2(z)={
{\prod\limits_{q}}^{(b)}\e^{\gamma(q)}
(1-\e^{-\gamma(q)}z)(1-\e^{-\gamma(q)}z^{-1}) \over
{\prod\limits_{q}}^{(f)}\e^{\gamma(q)}
(1-\e^{-\gamma(q)}z)(1-\e^{-\gamma(q)}z^{-1})
}.
\ee
Owing to the momentum  $q$ runs the entire Brillouin zone all the
factors of the r.h.s. (3.3) are double except for the values
 $q=0$ and  $q=\pi$ in the nominator. So the functions  $A(z)$,
$P(z)$ and $Q(z^{-1})$ have no branch points rather the simple poles
and zeroes:  \be P(z)= \exp\biggl({\gamma_0-\gamma_\pi\over2}\biggr)
{{\prod\limits_{0<q\leq\pi}}^{\!\!\!\!\!\!(b)}(\e^{\gamma(q)}-z)
\over{\prod\limits_{0<q<\pi}}^{\op (f)}(\e^{\gamma(q)}-z)},
\quad
Q(z^{-1})={{\prod\limits_{0\leq
q<\pi}}^{\op (b)}(1-\e^{-\gamma(q)}z^{-1})
\over{\prod\limits_{0<q<\pi}}^{\op (f)}(1-\e^{-\gamma(q)}z^{-1})}. \ee

According to (A25), (A28)
\be
\det A^{(r)}=\e^{h(r)}\sum_{l=0}^{\infty}g_{2l}(r).
\ee
Let us write the function  $h(r)$  as the sum of three
terms \be h(r)=-r/\Lambda+\ln \xi+\ln\xi_T.
\ee
It follows from the eqs. (A23) and (2.15)
\be \Lambda^{-1}=-[\ln P(0)+\ln Q(0)]=-{1\over2\pi
i}\oint\limits_{|z|=1}\!\!\!{dz\over z}\ln T(z)=
{1\over\pi}\int\limits_{0}^{\pi}dp\,\ln\coth({N\gamma(p)/2}).
\ee
The parameter $\Lambda$ has the sense of the coherence length. Its
asymptotic behaviour at large  $N$
 is \be \Lambda\simeq\e^{N\gamma_0}\sqrt{{\pi
N\over2\sinh \gamma_0}}.  \ee
One can see from  (3.8) that the coherence
length grows very rapidly when the cylinder circumference increases and
 $\gamma_0\neq0$. The lattice can be considered as infinite at
 $N\gamma_0\gg\max[\ln(N\gamma_0),\ln(r\gamma_0)]$. It follows also
from (A23) that  \be \ln(\xi\xi_T)= {1\over2\pi
i}\!\!\!\oint\limits_{|z|=1}\!\!\!\!dz\ln
Q(z^{-1}){\partial\over\partial z}\ln P(z)= {1\over(2\pi
i)^2}\!\!\!\oint\limits_{|z_2|>|z_1|}\!\!\!\!\!\!d(z_1z_2)\, {\ln
A(z_1)\ln A(z_2)\over(z_2-z_1)^2}, \ee
or separately for the $\ln\xi$ and $\ln\xi_T$
\be
\ln\xi={1\over4}\ln\biggl[{\sinh\gamma_0\cdot\sinh\gamma_\pi\over
\sinh^2\bigl((\gamma_0+\gamma_\pi)/2\bigr)}\biggr], \ee
\be
\ln\xi_T={1\over(2\pi
i)^2}\!\!\!\oint\limits_{|z_2|>|z_1|}\!\!\!\!\!\!{d(z_1z_2)\over(z_2-z_1)^2}\ln T(z_1)\ln
T(z_2).
\ee
The r.h.s. of the eq. (3.10) does not depend on $N$ and is singular at
the critical point \be
\ln\xi\simeq{1\over4}\ln\mu\quad \mbox{at}\quad \mu\to0.  \ee For the
calculation $\ln\xi_T$ let us
return to the momentum variables and
integrate  (3.11) by parts.
The result is
$$
\ln\xi_T={N^2\over2\pi^2}\int\limits_{0}^{\pi}{dp\, dq\,
\gamma'(p)\gamma'(q)\over\sinh (N\gamma(p))\sinh
(N\gamma(q))}\ln\biggl| {\sin((p+q)/2)\over\sin((p-q)/2)}\biggr| $$
or
\be
\ln\xi_T={N^2\over2\pi^2}\int\limits_{\gamma_0}^{\gamma_\pi}{d\gamma_p\, d\gamma_q\over\sinh (N\gamma(p))\sinh
(N\gamma(q))}\ln\biggl|{\sin((p+q)/2)\over\sin((p-q)/2)}\biggr|.
\ee
Differentiating eq. (3.13) with respect to $\mu$ one can obtain
\be
{\partial\over\partial\mu^2}\ln\xi_T=-{1\over2}[c_1^2(\mu)-c_2^2(\mu)],
\ee
where
\bea
&&c_1(\mu)={N\over2\pi}\int\limits_{\gamma_0}^{\gamma_\pi}{d\gamma\over\sinh(N\gamma)}
\sqrt{{\cosh\gamma_\pi-\cosh\gamma\over\cosh\gamma-\cosh\gamma_0}},
\\
&&c_2(\mu)={N\over2\pi}\int\limits_{\gamma_0}^{\gamma_\pi}{d\gamma\over\sinh(N\gamma)}
\sqrt{{\cosh\gamma-\cosh\gamma_0\over\cosh\gamma_\pi-\cosh\gamma}}.
\nonumber
\eea
Deriving eqs. (3.14), (3.15) one has to account for that the integrand
of  (3.13) vanishes on the boundaries of the integration region. Note
also that the derivatives of $p$ and $q$ are
$$ {\partial
p\over\partial\mu^2}=-{1\over2\sin p},\quad {\partial
q\over\partial\mu^2}=-{1\over2\sin q}, $$
that follows from  (2.18). For
$N\gamma_0\gg1$ the asymptotics of  $c_1(\mu)$ and $c_2(\mu)$ are
\be
c_1(\mu)\simeq{1\over\cosh(N\gamma_0)}\sqrt{{N\over2\pi\sinh\gamma_0}},
\quad c_2(\mu)\simeq{1\over4\cosh(N\gamma_0)}\sqrt{{\sinh
\gamma_0\over2\pi N}}.
\ee
Collecting  (3.14)--(3.16)  one can obtain
$$
\ln\xi_T=\int\limits_{\mu}^{\infty}d\nu\, \nu[c_1^2(\nu)-c_2^2(\nu)]
\simeq{1\over\pi}\e^{-2N\gamma_0}.
$$
So, we see that $\xi_T\to1$ if $N\to\infty$.

The integral (3.13) diverges at the critical point. This means that
$\ln\xi_T$ as a function of $\mu$ has a singularity. It easy to
obtain at $\mu\to0$
$$ c_1(\mu)\simeq1/2\mu,\quad
{\partial\ln\xi_T\over\partial\mu^2}=-{1\over8\mu^2} $$ and
consequently
\be \ln\xi_T=-{1\over4}\ln\mu+\const.  \ee
Therefore,  one can see from  (3.12), (3.17) that the function
$\ln(\xi\xi_T)$ has no singularity at $\mu=0$ if $N$ is finite.

Let now evaluate the expansion in the r.h.s. of (3.5)
\be
G_F(r)=\sum_{l=0}^{\infty}g_{2l}(r).
\ee
The coefficients $g_{2l}(r)$ are expressed through the
 $2l$-multiple contour integrals  (A30) as it is shown in the Appendix
\be g_{2l}(r)={(-1)^l\over l!l!(2\pi i)^{2l}}\oint\limits_{|z_i|<1}
{\prod\limits_{i=1}^{2l}(dz_i\,z_i^r)\prod\limits_{i=1}^{l-1}\prod\limits_{j=i+1}^{l}\bigl[
(z_{2i-1}-z_{2j-1})^2
(z_{2i}-z_{2j})^2\bigr]
\over\prod\limits_{i=1}^{l}\prod\limits_{j=1}^{l}(1-z_{2i-1}z_{2j})^2
}\
{\prod\limits_{i=1}^{l}W(z_{2i-1})\over\prod\limits_{i=1}^{l}W(z^{-1}_{2i})}.
\ee
The integral (3.19) is defined by the concrete form of the function
$W(z)=P(z)/Q(z^{-1})$: in our case this is the eq. (3.4).
The integrand in the r.h.s. of (3.19) is analytic at $|z_i|<1$ except
for the singularities which the functions  $W(z_i)$ and
$W^{-1}(z_i^{-1})$ possess.
To separate explicitly the singularities inside the circle  $|z|=1$
these functions ought to be rewritten in the following way
\be W(z)={P(z)\over Q(z^{-1})}=\sqrt{{(1-\e^{-\gamma_\pi}z)
(1-\e^{-\gamma_\pi}z^{-1})\over(1-\e^{-\gamma_0}z)(1-\e^{-\gamma_0}z^{-1})}}
{P^2_T(z)\over T(z)},
\ee
\be
W^{-1}(z^{-1})={Q(z)\over P(z^{-1})}=\sqrt{{(1-\e^{-\gamma_0}z)
(1-\e^{-\gamma_0}z^{-1})\over(1-\e^{-\gamma_\pi}z)(1-\e^{-\gamma_\pi}z^{-1})}}
{Q^2_T(z)\over T(z)}.
\ee
The analytic at $|z|<1$ functions $P_T(z)$ and $Q_T(z)$ are defined by
the  equations \be
\ln P_T(z)=\Pp\ln T(z)={1\over2\pi
i}\!\!\!\oint\limits_{|z|=1}\!\!\!\!{dz'\over
z'-z}\ln\tanh(N\gamma(p')/2), \ee \be \ln Q_T(z^{-1})=\Pm\ln
T(z)={1\over2\pi i}\!\!\!\oint\limits_{|z|=1}\!\!\!\!{dz'\over
z-z'}\ln\tanh(N\gamma(p')/2),  \ee
$$
z'=\e^{-ip'}.
$$
  Note that it follows from (3.7), (3.22),
(3.23)
$$ \ln P_T(0)=-1/\Lambda,\quad \ln Q_T(0)=0,\quad
P_T(z)=P_T(0)\cdot Q_T(z).  $$
With allowance of these eqs. one can obtain instead of
(3.20), (3.21)
\be W(z)=\e^{(\gamma_0-\gamma_\pi)/2-2/\Lambda}(\cosh
\gamma_\pi-\cos p)Q_T^2(z){\coth(N\gamma(p)/2)\over\sinh\gamma(p)},
\ee
\be
W^{-1}(z^{-1})=\e^{(\gamma_\pi-\gamma_0)/2}(\cosh
\gamma_0-\cos p)Q_T^2(z){\coth(N\gamma(p)/2)\over\sinh\gamma(p)},
\ee
\be
{\coth(N\gamma(p)/2)\over\sinh\gamma(p)}={1\over
N}{\sum_{q}}^{(b)}{1\over \cosh\gamma(q)-\cos p}.
\ee
It is seen from (3.24)--(3.26) that the singularities inside the
integration contours of (3.19) are the simple poles at the points
$$ z_{(k)}=\e^{-\gamma(q_{(k)})},\quad q_{(k)}={2\pi k\over N}, \quad
k=1,\ldots,N, $$ corresponding to the boson spectrum of the
quasimomentum $q$.  Accounting for $$ {1\over2\pi
i}\!\!\!\oint\limits_{|z|=1}\!\!\!\!{dz\over
z}\,{v(z)\over\cosh\gamma(q)-\cos
p}={v(\e^{-\gamma(q)})\over\sinh\gamma(q)}, $$ where $v(z)$
is analytic at $|z|<1$, the integral (3.19) is fulfilled by residua.
The result is
\be
g_{2l}(r)={\e^{-2l/\Lambda}\over(2l)!N^{2l}}{\sum\limits_{[q]}}^{(b)}
\prod\limits_{i=1}^{2l}
\biggl({\e^{-r\gamma_i-\eta_i}\over\sinh
\gamma_i}\biggr)\G_{2l}[q],
\ee
where
\bea
\G_{2l}[q]&=&C_l^{2l}
\prod\limits_{i=1}^{l-1}\prod\limits_{j=i+1}^{l}
[(\e^{-\gamma_{2i-1}}-\e^{-\gamma_{2j-1}})^2
(\e^{-\gamma_{2i}}-\e^{-\gamma_{2j}})^2]
\times
\nonumber\\
&&\times{
\prod\limits_{i=1}^{l}[(1+\cos
  q_{2i-1})(1-\cos q_{2i})]\over
  \prod\limits_{i=1}^{2l}\e^{\gamma_i}\prod\limits_{i=1}^{l}
  \prod\limits_{j=1}^{l}(1-
  \e^{-\gamma_i-\gamma_j})^2},
  \eea
  \be
  \gamma_i=\gamma(q_i),\quad \eta_i=\eta(q_i)=-\ln
  Q_T^2(\e^{-\gamma(q_i)}),
\ee
$$
{\sum_{[q]}}^{(b)}=
{\sum_{q_1}}^{(b)}{\sum_{q_2}}^{(b)}\cdots{\sum_{q_{2l}}}^{(b)},
$$
$$
C_l^{2l}={(2l)!\over l!l!}\quad-\quad\mbox{is the binomial
coefficient}. $$ Using the equalities
$$ \Pd(\e^{-\gamma_i}-\e^{-\gamma_j})(1-\e^{\gamma_i+\gamma_j})=\cosh
      \gamma_i -\cosh\gamma_j=\cos q_j-\cos q_i, $$
       the eq. (3.28) can be transformed to
       $$
\G_{2l}[q]={V_{2l}[q]\over\prod\limits_{i<j}^{2l}\sinh^2
((\gamma_i+\gamma_j)/2)},
$$
where
\bea
V_{2l}[q]&=&2^{-2l^2}C_l^{2l}\prod_{i=1}^{l-1}\prod_{j=i+1}^{l}[
(\cos q_{2i-1}-\cos q_{2j-1})^2(\cos q_{2i}-\cos q_{2j})^2]\times
\nonumber\\
&&\times\prod_{i=1}^{l}(1+\cos q_{2i-1})(1-\cos q_{2i}).
\eea
The expression (3.30) is not symmetric with respect to changes the
summation variables $q_i$ with even subscripts by that with odd
subscripts. The summation extracts the symmetric part of the
function  $V_{2l}[q]$ which is denoted by $V_{2l}^{\rm sim}[q]$
$$
V_{2l}^{\rm sim}[q]={1\over C_l^{2l}}\sum_{\rm P}V_{2l}[q],
$$
where
the sum over all permutations of $q_{2i-1}$ and $q_{2j}$ (theirs
number is just $C_l^{2l}$) is denoted by $\sum\limits_{\rm P}$\,. The
equality \be V_{2l}^{\rm sim}[q]=U_{2l}^{\rm even}[q], \ee is the
crucial for the final result.  Here the even (with respect to
$q_i\leftrightarrow-q_i$) part of the product $$
U_{2l}[q]=\prod_{i<j}^{2l}\sin^2\bigg({q_i-q_j\over2}\biggr), $$ is
denoted by $U_{2l}^{\rm even}[q]$
$$ U_{2l}^{\rm
even}[q]=2^{-2l}\sum_{[\sigma=\pm1]}\prod_{i<j}^{2l}\sin^2
\biggl({\sigma_iq_i-\sigma_jq_j\over2}\biggr)  $$
Consequently the function  $V_{2l}[q]$ under summation
(3.27) can be replaced by the function
$U_{2l}[q]$. This makes it possible to present the coefficient
$g_{2l}(r)$ in the form similar to that of eq. (1.2)
 $$ n=2l, $$ \be g_n(r)={\e^{-n/\Lambda}\over n!
N^n}{\sum_{[q]}}^{(b)}\prod_{i=1}^n
\biggl({\e^{-r\gamma_i-\eta_i}\over\sinh\gamma_i}\biggr)F_n^2[q_i],
\ee
where $F_n[q_i]$ is the lattice analog of the form factor (1.3)
\be
F_n[q_i]=\prod_{i<j}^{n}\left[{\sin((q_i-q_j)/2)\over
\sinh((\gamma_i+\gamma_j)/2)}\right],\quad F_0[q]=1.
\ee
I emphasize once more that the momentum spectrum over which
the summation (3.32) is extended (the intermediate states
summing up) occurs to be boson one contrary to the fermion
spectrum by which the initial values were defined, in particular
the matrix $A_{x,x'}$ (2.10).

The function  $\eta_i=\eta(q_i)$ that is contained in (3.32),
decreases rapidly when the cylinder circumference grows. It follows from
its definition (3.29)  \be \eta(q)=-{1\over \pi
i}\oint\limits_{|z|=1}{dz\ln
T(z)\over\e^{\gamma(q)}-z}={1\over\pi}\int\limits_{0}^{\pi}{dp(1-
\e^{-\gamma(q)}\cos p)\over\cosh\gamma(q)-\cos p}\ln\coth
(N\gamma(p)/2).  \ee One can find at $N\gamma_0\gg1$
$$ \eta(q)\simeq{4\over\e^{\gamma(q)}-1}{\e^{-N\gamma_0}
\over\sqrt{2\pi N\sinh\gamma_0}}\,,  $$
and $\eta(q)\to0$ if $N\to \infty$.

\renewcommand{\theequation}{4.\arabic{equation}}
\setcounter{equation}{0}

\section
{The Paramagnetic region: ${\bf K<K_c}$}

The correlation function is defined uniform through the determinant
$\det A^{(r)}$ in the whole region of parameters (including $K<K_c$).
But the method of its calculation is to be modified because the
Wiener--Hopf sum equation technique, which is used, demands
 the Toeplitz matrix kernel of the appropriate form. Both the kernel
and the logarithm of the kernel have to satisfy the Loran expansion
condition.  Meanwhile, the matrix kernel (2.10) is reforming compare
 to the ferromagnetic case (2.15), when $K$ goes across $K_c$ $$
 A(z)=z^{-1}B(z), $$ where \be
B(z)=\sqrt{{(1-\e^{-\gamma_0}z)(1-\e^{-\gamma_\pi}z)
\over(1-\e^{-\gamma_0}z^{-1})(1-\e^{-\gamma_\pi}z^{-1})}}T(z), \ee
so, rather the function $B(z)$ than $A(z)$ possesses appropriate
analytical properties. The matrix  $A^{(r)}$ now has the following
form \be A_{x,x'}= -{1\over2\pi
i}\oint\limits_{|z|=1}{dz\over z}z^{-x+x'-1}B(z).  \ee Factorizing the
kernel (4.1)
 $$ B(z)=P(z)\cdot Q(z^{-1}), $$
where the functions $P(z)$ and $Q(z)$ are analytic inside the circle
$|z|<\e^{\gamma_0}$ one can obtain
\be
P(z)=\exp\biggl(-{\gamma_0+\gamma_\pi\over2}\biggr){
{\prod\limits_{0\leq q\leq\pi}}^{\op(b)}(\e^{\gamma(q)}-z)
\over
{\prod\limits_{0<q<\pi}}^{\op(f)}(\e^{\gamma(q)}-z)
},\qquad
Q(z^{-1})={
{\prod\limits_{0<q<\pi}}^{\op(b)}(1-\e^{-\gamma(q)}z^{-1})
\over
{\prod\limits_{0<q<\pi}}^{\op(f)}(1-\e^{-\gamma(q)}z^{-1})
}.\ee
Comparing with the factorized representation in the ferromagnetic case
one can see that the corresponding products
(3.4) and (4.3) differ from each other by one term:
the factor $(\e^{\gamma_0}-z)$ is appeared in
 $P(z)$ and the factor
$(1-\e^{-\gamma_0}z^{-1})$ is disappeared in
 $Q(z^{-1})$. It follows from
 (4.3)
$$ P(0)=P_T(0)=\e^{-1/\Lambda},\qquad Q(0)=Q_T(0)=1,
$$ where the functions  $P_T(z)$ and $Q_T(z)$ are the same as in the
ferromagnetic case (3.22), (3.23).

The matrix (4.2) has the structure (A31) considered in the Appendix.
Therefore its determinant can be represented according to
the eqs. (A32)--(A35) by the following way
\be \det
A^{(r)}=\e^{h(r+1)}F(r), \ee $$ F(r)=\sum_{l=0}^{\infty}f_{2l+1}(r).
$$
The function  $h(r+1)$ is given by eq. (A23). Writing it similarly to
 (3.6) and allowing for the difference between the definitions  (4.3)
and (3.4) for the corresponding functions $P(z)$ and $Q(z)$ one can
obtain
\be
h(r+1)=-(r+1)/\Lambda+\ln(\xi\xi_T)+\ln(1-\e^{-\gamma_0-\gamma_\pi}),
\ee where the values $\Lambda$, $\xi$ and $\xi_T$ are the same as in
the ferromagnetic phase (3.7),  (3.10) and (3.11). With the
allowance (4.5) the eq. (4.4) can be rewritten as  follows
\be\det
A^{(r)}=\sinh((\gamma_0+\gamma_\pi)/2)\e^{h(r)}G_P(r), \ee where \be
G_P(r)=2\e^{-1/\Lambda-(\gamma_0+\gamma_\pi)/2}F(r)=\sum_{l=0}^{\infty}
g_{2l+1}(r),
\ee
and for the coefficients $g_{2l+1}(r)$ in the expansion  (4.7) one
obtains from (A35) \bea && \ \hspace{-2cm}
g_{2l+1}(r)=2\e^{-1/\Lambda-(\gamma_0+\gamma_\pi)/2} {(-1)^l\over
l!(l+1)!}(2\pi
i)^{2l+1}\oint\limits_{|z_i|<1}\prod_{i=1}^{2l+1}(dz_i\, z_i^r)\times
\nonumber\\
&&
\ \hspace{-2cm}
\hphantom{f_{2l+1}(r)=}
\times
{
\prod\limits_{i=1}^{l-1}\prod\limits_{j=i+1}^{l}(z_{2i}-z_{2j})^2
\prod\limits_{i=1}^{l}\prod\limits_{j=i+1}^{l+1}(z_{2i-1}-z_{2j-1})^2
\over
\prod\limits_{i=1}^{l}\prod\limits_{j=0}^{l}(1-z_{2i}z_{2j+1})^2
}\
{
\prod\limits_{i=1}^{l}(z_{2i}W(z_{2i}))
\over
\prod\limits_{i=0}^{l}(z_{2i+1}W(z_{2i+1}^{-1}))
}\,.
\eea
The functions $W(z)=P(z)/Q(z^{-1})$ and
 $W^{-1}(z^{-1})=Q(z)/P(z^{-1})$ entering the integrals (4.8) have the
following form in this case \bea
W(z)&=&2\e^{-2/\Lambda-(\gamma_0+\gamma_\pi)/2}Q_T^2(z)\sinh^2\gamma(p)
{1\over N}{\sum_{q}}^{(b)}{1\over\cosh \gamma(q)-\cos p},
\\
W^{-1}(z^{-1})&=&{1\over2}\e^{(\gamma_0+\gamma_\pi)/2}Q_T^2(z)
{1\over N}{\sum_{q}}^{(b)}{1\over\cosh \gamma(q)-\cos p}.
\eea
It follows from (4.9), (4.10) that the integrand in
(4.8) has only the simple poles inside the integration contour at the
points $$
z_{(k)}=\e^{-\gamma(q_{(k)})},\quad q_{(k)}={2\pi k\over N},\quad
k=1,2,\ldots, N; $$ similar to the previous case. At these points,
 obviously, $$ \cos p=\cosh\gamma(q),\quad \cosh \gamma(p)=\cos
q,\quad\sinh^2\gamma(p)=-\sin^2q.  $$
So, the integration contours in (4.8) may be squeezed and the
coefficient $g_{2l+1}(r)$ is expressed through the sum of residua \be
g_{2l+1}(r)={\e^{-(2l+1)/\Lambda}\over(2l+1)!N^{2l+1}}{\sum_{[q]}}^{(b)}
\prod_{i=1}^{2l+1}\biggl({\e^{-r\gamma_i-\eta_i}\over\sinh\gamma_i}\biggr)
{V_{2l+1}[q]\over\prod\limits_{i<j}^{2l+1}\sinh^2((\gamma_i+\gamma_j)/2)},
\ee
where
\bea
V_{2l+1}[q]&=&C_l^{2l+1}2^{-2l(l+1)}
\prod_{i=1}^{l-1}\prod_{j=i+1}^{l}
(\cos q_{2i}-\cos q_{2j})^2\times\nonumber\\
&&\times\prod_{i=1}^{l}\prod_{j=i+1}^{l+1}
(\cos q_{2i-1}-\cos q_{2j-1})^2
\prod_{i=1}^{l}\sin^2q_{2i}.
\eea
The expression (4.12) differs from the corresponding ferromagnetic one
(3.30). But it occurs that also in the paramagnetic case
the symmetric part of the function
$V_{2l+1}[q]$ -- with respect to change of each variable with even
subscript $q_{2i}$ by each variable with odd subscript
 $q_{2j-1}$ (the number of the permutations is $C_l^{2l+1}$) --
coincides with the even part of the function $U_{2l+1}[q]$
$$
U_{2l+1}[q]=\prod_{i<j}^{2l+1}\sin^2\biggl({q_i-q_j\over2}\biggr). $$
The equality analogous to the eq. (3.31) takes place
$$ V_{2l+1}^{\rm sim}[q]=U_{2l+1}^{\rm even}[q], $$ where $$
V_{2l+1}^{\rm sim}[q]={1\over C_l^{2l+1}}\sum_{\rm P}V_{2l+1}[q], $$
$$ U_{2l+1}^{\rm
even}[q]=2^{-(2l+1)}\sum_{[\sigma=\pm1]}\prod_{i<j}^{2l+1}
\sin^2\biggl({\sigma_iq_i-\sigma_jq_j\over2}\biggr).
$$
Consequently the function $V_{2l+1}[q]$ can be replaced by the
 function $U_{2l+1}[q]$  under the summation in the r.h.s. of eq.
   (4.11). We obtain eventually
 $$
n=2l+1, $$
\bea g_n(r)&=&{\e^{-n/\Lambda}\over
n!N^n}{\sum_{[q]}}^{(b)}\prod_{i=1}^{n}
\biggl({\e^{-r\gamma_i-\eta_i}\over\sinh\gamma_i}\biggr)F^2_n[q],\\
F_n[q]&=&\prod_{i<j}^{n}{\sin((q_i-q_j)/2)\over\sinh((\gamma_i+\gamma_j)/2)},
\quad F_1[q]=1.
\eea
The functions $\gamma_i=\gamma(q_i)$ and $\eta_i=\eta(q_i)$
are defined by the eq. (3.29).

The form of the coefficients $g_n(r)$ and of the function $F_n[q]$
is surprisingly the same both the ferro- and paramagnetic cases. In
the ferromagnetic case the correlation function expansion is extended
over even $n$ and for the paramagnetic case -- over odd $n$
\bea
\langle\sigma{(\rv_1)}\sigma{(\rv_2)}\rangle&=&(\xi\cdot\xi_T)\e^{-r/\Lambda}
\sum_{l=0}^{\infty}g_{2l}(r),\qquad K>K_c;\\
\langle\sigma{(\rv_1)}\sigma{(\rv_2)}\rangle&=&
\sinh\biggl({\gamma_0+\gamma_\pi\over2}\biggr)
(\xi\cdot\xi_T)\e^{-r/\Lambda}
\sum_{l=0}^{\infty}g_{2l+1}(r),\quad K<K_c.
\eea

Notice that the even values of the circumference $N$ were assumed in the
factorization procedure for the functions $P(z)$ and $Q(z^{-1})$
(see eqs. (3.4) and (4.3)\ ). For the odd $N$ the Brillouin zone does
not contain the
 $q=\pi$ point in the boson case. On the contrary this value appears
in the fermion spectrum. It is not difficult to account for this
detail, it is irrelevant for the final result.

\renewcommand{\theequation}{5.\arabic{equation}}
\setcounter{equation}{0}

\section
{Conclusion}

It should be observed first of all that, although the
representations (4.15), (4.16) for the correlation function are
obtained  for  different regions  of the coupling parameter ($K>K_c$
and $K<K_c$ respectively), both of them are valid and equal to each
other for all values of $K$ at any finite cylinder circumference  $N$.
 For the finite $N$ the expansions (4.15), (4.16) are not series
 rather finite sums due to the form factor $F_n[q]$ is identically
 zero for $n>N$. Only the thermodynamic limit $N\to\infty$ splits the
 domain of their validity.

The obtained lattice form factor representations possess
one beautiful feature:  they can be checked numerically or
analytically.  For small values of $N$ the result may be computed in
other way, for example with the help of the transfer matrix
technique. In particular at $N=1$ we have the one dimensional Ising
chain.  The elementary calculation gives \be
\langle\sigma(0)\sigma(r)\rangle=t^r.  \ee One can see after simple
transformations that both the representations (4.15) and (4.16)
coincide with (5.1).

Apart from its intrinsic interest the Ising model deserves scrutiny
because of its relevance to nonperturbative explorations in quantum
field theory. Mutual correspondence between field model and spin
lattice system in scaling regime implies in our  case the
limit
$K\to K_c$,\ \
$(N,M,r)\to\infty$\ \  at \ \
$(N\mu,M\mu,r\mu)=\const$.
Mysterious process of thermodynamic or scaling
limit may be observed in detail proceeding from the explicit
expressions obtained in this work.

The IM correlation function
$\langle\sigma(\rv)\sigma(\rv')\rangle$
in scaling limit corresponds to the quantum field model two point
Green function $G(x,t;L,\beta)$. The variables of the Green
function are assumed to be scaled on mass:  $m=1$, $x=\mu|r_x-r'_x|$
 is the spatial distance, $t=\mu|r_y-r'_y|$ is the temporal one.
$L=\mu M$ is the volume, $\beta=\mu N$ is the temperature. The
boundary conditions along the temporal coordinate are quite definite:
periodic for the boson field, antiperiodic for the fermion field. This
is the issue of the appropriate formulation of the quantum field
theory at finite temperature. On the contrary, the spatial boundary
condition may be arbitrary. In any case the presence of boundaries
breaks down explicitly the Lorentz invariance (the rotational
invariance in Euclidian metric case). The spatial and temporal
coordinates are not equivalent in general.  In this sense the
configuration of correlating spins considered in this paper
corresponds to the equal time Green function
$G(x,0;\infty,\beta)$ in the infinite volume at finite
temperature. With some stipulations this function may be viewed also
as that at zero temperature in finite volume $\widetilde{L}=\mu N$
(with periodicity in spatial coordinate). One has to keep in mind in
addition that this is not equal time rather equal site Green function
$G(0,\widetilde{t};\widetilde{L},\infty)$.
It is necessary to consider the correlating spins placed on a cylinder
circumference to compute the equal time Green function in finite
volume and zero temperature. The corresponding problem may be solved
by the method used in this work but with the Toeplitz determinant
  technique being properly modified.

Meanwhile, the representation (4.13)--(4.16) for the IM correlation
  function has such transparent structure that the conjecture for
  the case of general displacement  of correlating spins suggests
  itself.  The simplest one is \bea
\langle\sigma(0)\sigma(r_x,r_y)\rangle&=&(\xi\cdot\xi_T)\e^{-r_x/\Lambda}
\sum_{n}g_{n}(r_x,r_y),\\
 g_n(r_x,r_y)&=&{\e^{-n/\Lambda}\over
n!N^n}{\sum_{[q]}}^{(b)}\prod_{j=1}^{n}
\biggl({\e^{-r_x\gamma_j-ir_yq_j-\eta_j}\over\sinh\gamma_j}\biggr)F^2_n[q].
\eea
It is amazing that this generalization is in exact agreement with the
transfer matrix results for the N-rows Ising chains. We have checked
it analytically for $N=2,3,4$ and numerically for $N=5,6$.

\bigskip
I thank V.Shadura for the fruitful discussions and O.Lisovy for the
assistance in  computations. I am grateful to Prof. B.M.McCoy for his
interest to my work and useful comments.

This work was supported by the INTAS Program (Grant
INTAS-97-1312).

\renewcommand{\theequation}{A\arabic{equation}}
\setcounter{equation}{0}

\vspace{1cm}
\hfill {\Large\bf Appendix}

\medskip
The elements of the Toeplitz matrix arranged along the diagonals
are equal \be A_{x,x'}=A_k \ \ \mbox{for}\ \
k=x-x'=0,\,\pm1,\,\ldots,\,\pm r-1\,, \ee \be A^{(r)}=
\left(\begin{array}{ccccc} A_0&A_{-1}&A_{-2}&\cdots&A_{-r+1}\\
A_1&A_{0}&A_{-1}&\cdots&A_{-r+2}\\
A_2&A_{1}&A_{0}&\cdots&A_{-r+3}\\
\vdots&\vdots&\vdots&\ddots&\vdots\\
A_{r-1}&A_{r-2}&A_{r-3}&\cdots&A_{0} \end{array}\right)\,. \ee The
superscript in $A^{(r)}$ shows the matrix dimension. If there is
the function of the complex variable $A(z)$ analytical in the ring
 $\alpha<|z|<\alpha^{-1}$ $(\alpha<1)$ such that the coefficients
$A_k$ of its Loran series \be A(z)=\sum_{k=-\infty}^{\infty}A_kz^k \ee
coincide with the matrix element  (A2) then the matrix  (A1) can be
represented in the form \be A_{x,x'}={1\over 2\pi
i}\!\!\oint\limits_{|z|=1}\!\!{dz\over z}z^{-x+x'}A(z).  \ee The
function $A(z)$ in the integrand of  (A4) is called the kernel of the
Toeplitz matrix $A^{(r)}$.

If the inverse matrix is known then the calculation of the determinant
reduces to calculation of trace with help of the identity
\be
\ln \det A^{(r)}(\lambda)=\int d\lambda\,
\Sp\biggl[\bigl(A^{(r)}(\lambda)\bigr)^{-1}{\partial\over\partial
\lambda} A^{(r)}(\lambda)\biggr]+\const, \ee
where $\lambda$ is a parameter of which the matrix elements depend on
explicitly.

The inversion procedure for the Toeplitz matrix simplifies
if both the kernel
(A3) and  $\ln A(z)$ may be expanded in Loran series. If so, the
kernel $A(z)$ can be expressed in factorized form
\be A(z)=P(z)\cdot
Q(z^{-1}). \ee
The functions $P(z)$ and  $Q(z)$ are analytic inside the circle
$|z|\leq\alpha^{-1}$: \be \ln P(z)=\Pp\ln A(z), \quad \ln
Q(z^{-1})=\Pm\ln A(z).  \ee The projection operators $\Pp$ and  $\Pm$
extract the analytic part of the given function $v(z)$ inside and
outside the circle $|z|=1$ correspondingly $$ \Pp^2=\Pp,\quad
\Pm^2=\Pm,\quad \Pp\Pm=\Pm\Pp=0,\quad \Pp +\Pm =1, $$ \be \Pp
v(z)={1\over2\pi i}\!\!\!\oint\limits_{|z|=1}\!\!\!{d\xi v(\xi)\over
\xi-z}, \quad \Pm v(z)={1\over2\pi
i}\!\!\!\oint\limits_{|z|=1}\!\!\!{d\xi v(\xi)\over z- \xi}.  \ee

The problem of the Toeplitz matrix inversion can be reduced to solving
the system of the Wienner-Hopf sum equations [3]. In terms of the
projection operators
 (A8)
the matrix elements of the inverse matrix
$(A^{(r)})^{-1}_{x,x'}$ are
\bea && \hspace{-1cm}
(A^{(r)})^{-1}_{x,x'}=\Pp{z^{r-x}\over Q(z^{-1})}\Pm w^{-1}(z)
\bigl[1-\Pm w(z)\Pp w^{-1}(z)\bigr]^{-1}\Pp{z^{x'}\over Q(z^{-1})}
\biggm|_{z=0}\,,\\ && \hspace{-1cm}
(A^{(r)})^{-1}_{x,x'}=\Pp{z^{-x}\over P(z)}\Pp w(z) \bigl[1-\Pp
w^{-1}(z)\Pm w(z)\bigr]^{-1}\Pm {z^{x'-r}\over P(z)} \biggm|_{z=0}\,,
\\ &&
\hspace{-1cm} (A^{(r)})^{-1}_{x,x'}=\Pp{z^{-x}\over P(z)}
\biggl\{1-\Pp\bigl[1-w(z)\Pp w^{-1}(z)\Pm\bigr]^{-1}w(z)\Pp
w^{-1}(z)\biggr\} \Pp{z^{x'}\over Q(z^{-1})} \biggm|_{z=0}\,, \eea
where \be w(z)=z^rW(z),\quad W(z)={P(z)\over Q(z^{-1})}.  \ee Each
of the alternative but equivalent representations
 (A9)--(A11) for the inverse matrix $(A^{(r)})^{-1}_{x,x'}$ are
preferable for the calculation of different matrix elements. In
particular, the eq.~(A9) is suitable for the calculation of the
 element $(A^{(r)})^{-1}_{r-1,0}$, the eq. (A11) -- for the element
$(A^{(r)})^{-1}_{0,0}$:
\bea &&(A^{(r)})^{-1}_{r-1,0}={1\over
Q^2(0)}\biggl\{z\Pm w^{-1}(z)\bigl[1-\Pm w(z)\Pp
w^{-1}(z)\bigr]^{-1}\biggr\} \biggm|_{z\to \infty}\,,\\
&&(A^{(r)})^{-1}_{0,0}={1\over P(0)Q(0)}\biggl\{1-\Pp\bigl[1-w(z)\Pp
w^{-1}(z)\Pm\bigr]^{-1}w(z) \Pp w^{-1}(z)\biggr\}
\biggm|_{z=0}\,.  \eea
Expanding the inverse operators that are contained in the r.h.s.
of (A13) and  (A14) into geometrical progression $$ \bigl[1-\Pm
w(z)\Pp w^{-1}(z)\bigr]^{-1} =\sum_{l=0}^{\infty} \bigl[\Pm w(z)\Pp
w^{-1}(z)\bigr]^{l}, $$ $$ \bigl[1-w(z)\Pp w^{-1}(z)\Pm\bigr]^{-1}
=\sum_{l=0}^{\infty} \bigl[w(z)\Pp w^{-1}(z)\Pm\bigr]^{l} $$ and
using the integral representations for the projection operators
 $\Pp$, $\Pm$
(A8) one can obtain for (A13) with allowance (A12)
\be (A^{(r)})^{-1}_{r-1,0}={1\over
Q^2(0)}\sum_{l=0}^{\infty}a_{2l+1}(r), \ee \be
a_{2l+1}(r)={1\over(2\pi
i)^{2l+1}}\oint\limits_{|z_i|<1}{\prod\limits_{i=1}^{2l+1}(dz_i\,z_i^r)\over z_1z_{2l+1}\prod\limits_{i=1}^{2l}(1-z_iz_{i+1})}\
{\prod\limits_{i=1}^{l}W(z_{2i})\over\prod\limits_{i=0}^{l}W(z^{-1}_{2i+1})}.
\ee
In particular, the coefficients  $a_1(r)$ and $a_3(r)$ are
\bea
&&a_1(r)=
{1\over2\pi i}\oint\limits_{|z|=1}{dz\, z^{r-2}\over W(z^{-1})},
\nonumber\\
&&a_3(r)=
{1\over(2\pi i)^3}\oint\limits_{|z_i|<1}{d(z_1z_2z_3)\,
(z_1z_2z_3)^{r}\over z_1z_3(1-z_1z_2)(1-z_2z_3)}\ {W(z_2)\over
W(z^{-1}_1)W(z^{-1}_3)}.
\nonumber
\eea
Analogously one can  obtain for the eq. (A14)
\be (A^{(r)})^{-1}_{0,0}={1\over
P(0)Q(0)}\biggl(1-\sum_{l=1}^{\infty}a_{2l}(r)\biggr), \ee
\be
a_{2l}(r)={1\over(2\pi
  i)^{2l}}\oint\limits_{|z_i|<1}{\prod\limits_{i=1}^{2l}(dz_i\,
  z_i^r)\over z_1z_{2l} \prod\limits_{i=1}^{2l-1}(1-z_i
z_{i+1})}\ {\prod\limits_{i=1}^{l}W(z_{2i-1})
\over\prod\limits_{i=1}^{l}W(z_{2i}^{-1})}
\ee
and, in particular,
\bea
&&a_2(r)={1\over(2\pi
i)^2}\oint\limits_{|z_i|<1}{d(z_1z_2)(z_1z_2)^r\over
z_1z_2(1-z_1z_2)}\ {W(z_1)\over W(z^{-1}_2)},
\nonumber\\
&&a_4(r)={1\over(2\pi
i)^4}\oint\limits_{|z_i|<1}{d(z_1z_2z_3z_4)(z_1z_2z_3z_4)^r\over
z_1z_4(1-z_1z_2)(1-z_2z_3)(1-z_3z_4)}\ {W(z_1)W(z_3)\over
W(z^{-1}_2)W(z^{-1}_4)}.
\nonumber
\eea

 Consider now the auxiliary Toeplitz matrix $A^{(r)}(\lambda)$
depending on the parameter $\lambda$,
\be
A_{x,x'}(\lambda)={1\over2\pi i}\oint\limits_{|z|=1}{dz\over
z}z^{-x+x'}P(z)Q^\lambda(z^{-1}),\quad 0\leq\lambda\leq1,
\ee
where
the functions  $P(z)$ and $Q(z)$ are the same as in the kernel of the
matrix $A^{(r)}$. It is seen from the definition (A19) that at $\lambda=1$

$$ A_{x,x'}(1)={1\over2\pi
i}\oint\limits_{|z|=1}{dz\over z}z^{-x+x'}P(z)Q(z^{-1})=A_{x,x'}, $$
and at $\lambda=0$
\be
A_{x,x'}(0)={1\over2\pi i}\oint\limits_{|z|=1}{dz\over
z}z^{-x+x'}P(z).
\ee
The matrix (A20) is the triangular one: all its elements to the right
of the main diagonal are equal to zero $$ A_{x,x}(0)=P(0), \quad
A_{x,x'}(0)=0\ \ \mbox{if}\ \  x'>x. $$ Therefore \be \det
A^{(r)}(0)=P^r(0).  \ee The integration constant in (A5) is defined
with help of (A21).

Let us select the contribution $h(r)$ in  $\ln
 \det A^{(r)}$ which does not vanish when the matrix dimension grows
$$ \ln \det A^{(r)}=h(r)+\Delta h(r),\quad \Delta h(r)\to0\ \
\mbox{if}\ \ r\to \infty.  $$ It is not difficult to observe that
only the first term of the inverse operator expansion in (A10)
contributes in $h(r)$. This fact results in drastic simplification
of corresponding computations.  The inverse matrix (A10) is \bea
\bigl(A^{(r)}(\lambda)\bigr)^{-1}_{x,x'}&=&\Pp{z^{-x}\over
P(z)}\Pp {z^rP(z)\over Q^\lambda(z^{-1})}\Pm{z^{x'r}\over P(z)}=
\nonumber\\ &=&{1\over(2\pi i)^3}\oint\limits_{|z_2|>|z_{1,3}|}
{d(z_1z_2z_3)z_1^{-x}(z_2/z_3)z_3^{-x'}\over
z_1(z_2-z_1)(z_2-z_3)}\ {P(z_2)\over
P(z_1)Q^\lambda(z_2^{-1})P(z_3)}. \nonumber \eea The derivative of
the matrix
  (A19) is $$
{\partial\over\partial\lambda}A_{x,x'}^{(r)}(\lambda)={1\over2\pi
i}\oint\limits_{|z|=1}{dz\over z}z^{-x+x'}P(z)Q^\lambda(z^{-1})\ln
Q(z^{-1}), $$ and finally one has for the trace in (A5) \bea
&&\hspace{-.5cm}\Sp\biggl[\bigl(A^{(r)}(\lambda)\bigr)^{-1}
{\partial\over\partial\lambda} A^{(r)}(\lambda)\biggr]=
{1\over(2\pi
i)^4}\oint\limits_{|z_{2i}|>|z_{2j-1}|}\prod_{i=1}^{4}dz_i\times
\nonumber\\ &&\ \quad \times{
\bigl[1-(z_4/z_1)^r\bigr]\bigl[1-(z_3/z_4)^r\bigr](z_2/z_3)^r\over
(z_2-z_1)(z_2-z_3)(z_4-z_1)(z_4-z_3)}\cdot
{P(z_2)P(z_4)Q^\lambda(z_4^{-1})\ln Q(z_4^{-1})\over
P(z_1)Q^\lambda(z_2^{-1})P(z_3)} = \nonumber\\ &&\
\hspace{-.5cm}\quad= r\ln Q(0)+{1\over2\pi
i}\oint\limits_{|z|=1}dz\ln Q(z^{-1}){\partial\over\partial z}\ln
P(z)+o(r^{-1}). \eea It is seen from (A22) that nonvanishing
contribution in the trace occurs to be independent on the
parameter
 $\lambda$. Therefore, with allowance (A21) we obtain
 \be h(r)=r[\ln
P(0)+\ln Q(0)]+{1\over2\pi i}\oint\limits_{|z|=1} dz \ln
Q(z^{-1}){\partial\over\partial z}\ln P(z).  \ee For the
calculation of $\Delta h(r)$ the eq. (A5) may be used by keeping
the next terms in the expansion of (A10). But it is more
convenient for this purpose to exploit the specific property of
the Toeplitz matrix i.e. $$ \bigl(A^{(r+1)}\bigr)^{-1}_{0,0}={\det
A^{(r)}\over \det A^{(r+1)}}\,. $$ Therefore $$ \det A^{(r)}= \det
A^{(r+k)}\prod_{s=r+1}^{r+k}\bigl(A^{(s)}\bigr)^{-1}_{0,0},$$ and
after substitution (A17) we obtain \be \det A^{(r)}=
{\rm{e}}^{h(r+k)+\Delta h(r+k)}
[P(0)Q(0)]^{-k}\prod_{s=r+1}^{r+k}\biggl(1-\sum_{l=1}^{\infty}
a_{2l}(s)\biggr). \ee It follows from  (A23) that $$
{\rm{e}}^{h(r+k)}=[P(0)Q(0)]^{k}{\rm{e}}^{h(r)}, $$ so, the
factors $P(0)$ and $Q(0)$ in (A24) cancel. Due to this fact the
limit $k\to\infty$ in (A24) can be taken \be \det
A^{(r)}=G(r)\cdot\e^{h(r)}, \ee where \be
G(r)=\prod_{s=r+1}^{\infty}\biggl(1-\sum_{l=1}^{\infty}a_{2l}(s)\biggr).
\ee Expanding the product of sums in the r.h.s. of (A26)
\be
G(r)
=1-\sum_{s_1=r+1}^{\infty}
\biggl(\sum_{l=1}^{\infty}a_{2l}(s)\biggr)+
\sum_{s_1=r+1}^{\infty}
\sum_{s_2=s_1+1}^{\infty}
\biggl(\sum_{l=1}^{\infty}a_{2l}(s_1)\biggr)
\biggl(\sum_{l=1}^{\infty}a_{2l}(s_2)\biggr)-\cdots\, ,
\ee
summing up over $s_i$ and collecting the terms with the definite
multiplicity of integration, one can represent (A27) in the form
\be G(r)=\sum_{l=0}^{\infty}g_{2l}(r).  \ee The coefficients
$g_{2l}(r)$ can be obtained from (A27) directly but it is more
simple to compute them by use the following recursion relation
  \be
g_{2l}(r)+\sum_{k=0}^{l-1}\sum_{s=r+1}^{\infty}g_{2k}(s)a_{2(l-k)}(s)=0.
\ee
It is not difficult to deduce this relation from (A27).
Using the eq. (A18) for $a_{2l}(s)$ and initial condition
 $g_0(s)=1$ one can obtain from (A29)
 \be g_{2l}(r)={(-1)^l\over
l!l!(2\pi i)^{2l}}\oint\limits_{|z_i|<1}
{\prod\limits_{i=1}^{2l}(dz_i\,z_i^r)\prod\limits_{i=1}^{l-1}\prod\limits_{j=i+1}^{l}\bigl[ (z_{2i-1}-z_{2j-1})^2
(z_{2i}-z_{2j})^2\bigr]
\over\prod\limits_{i=1}^{l}\prod\limits_{j=1}^{l}(1-z_{2i-1}z_{2j})^2
}\
{\prod\limits_{i=1}^{l}W(z_{2i-1})\over\prod\limits_{i=1}^{l}W(z^{-1}_{2i})},
\ee
in particular,
\bea
&&g_2(r)=- {1\over(2\pi i)^{2}}\oint\limits_{|z_i|<1}{d(z_1z_2)\,
(z_1z_2)^r\over(1-z_1z_2)^2} {W(z_1)\over W(z_2^{-1})},
\nonumber\\
&&g_4(r)=- {1\over4(2\pi
i)^{4}}\oint\limits_{|z_i|<1}{d(z_1z_2z_3z_4)\,
(z_1z_2z_3z_4)^r(z_1-z_3)^2(z_2-z_4)^2\over(1-z_1z_2)^2
(1-z_2z_3)^2(1-z_3z_4)^2(1-z_4z_1)^2} {W(z_1)W(z_3)\over
                                    W(z_2^{-1})W(z_4^{-1})}.
\nonumber\eea
Note that the integrand of (A30) is symmetric both with respect to
permutations of the even subscript variables and with respect to
permutations odd subscript variables.

The structure of Toeplitz matrix is altered in paramagnetic region.
Here it is necessary to evaluate the determinant of the matrix which
has the following form \be
B^{(r)}=\left(\begin{array}{ccccc} A_{-1}&A_{0}&A_{1}&\cdots&A_{r-2}\\
A_{-2}&A_{-1}&A_{0}&\cdots&A_{r-3}\\
A_{-3}&A_{-2}&A_{-1}&\cdots&A_{r-4}\\
\vdots&\vdots&\vdots&\ddots&\vdots\\
A_{-r}&A_{-r+1}&A_{-r+2}&\cdots&A_{-1}
\end{array}\right),
\ee
$$
B_{x,x'}={1\over2\pi i}\oint\limits_{|z|=1}{dz\over z}
z^{-x+x'-1}A(z).
$$
The technique stated above is not valid because the kernel of the
matrix $B^{(r)}$ $$
B(z)=z^{-1}A(z) $$
does not satisfy the Loran expansion condition for $\ln B(z)$ if
 $\ln A(z)$ does satisfy. Nevertheless, the problem can be reduced to
that considered above. Really, it follows from the definition of the
inverse matrix element through the determinant and cofactor
$$
\det B^{(r)}=(-1)^r(A^{(r+1)})^{-1}_{r,0}\det A^{(r+1)}.  $$
Using the eqs. (A15), (A16) for the inverse matrix and the eqs.
(A25), (A28), (A30) for the determinant one can obtain \be \det
B^{(r)}=(-1)^r\e^{h(r+1)}F(r), \ee
  where \be F(r)=G(r+1)\cdot (A^{(r+1)})^{-1}_{r,0}={1\over
  Q^2(0)}\sum_{l=0}^{\infty}f_{2l+1}(r), \ee \be
f_{2l+1}(r)=\sum_{k=0}^{l}a_{2k+1}(r+1)g_{2(l-k)}(r+1).  \ee
All summands in the r.h.s. of
(A34) are the integrals of  $2l+1$ multiplicity. After reduction the
integrands to common denominator and symmetrization over even
subscript variables and over odd subscript variables one can obtain
 \bea
&& \ \hspace{-2cm} f_{2l+1}(r)={(-1)^l\over l!(l+1)!(2\pi
i)^{2l+1}}\oint\limits_{|z_i|<1}\prod_{i=1}^{2l+1}(dz_i\, z_i^r)\times
\nonumber\\
&&
\ \hspace{-2cm}
\hphantom{f_{2l+1}(r)=}
\times
{
\prod\limits_{i=1}^{l-1}\prod\limits_{j=i+1}^{l}(z_{2i}-z_{2j})^2
\prod\limits_{i=1}^{l}\prod\limits_{j=i+1}^{l+1}(z_{2i-1}-z_{2j-1})^2
\over
\prod\limits_{i=1}^{l}\prod\limits_{j=0}^{l}(1-z_{2i}z_{2j+1})^2
}\
{
\prod\limits_{i=1}^{l}(z_{2i}W(z_{2i}))
\over
\prod\limits_{i=0}^{l}(z_{2i+1}W(z_{2i+1}^{-1}))
},
\eea
and, in particular,
\bea
&&f_1(r)={1\over2\pi i}\oint\limits_{|z|=1}{dz\,
z^r\over zW(z^{-1})},
\nonumber\\
&&f_3(r)=- {1\over2(2\pi
i)^{3}}\oint\limits_{|z_i|<1}{d(z_1z_2z_3)\,
(z_1z_2z_3)^r(z_1-z_3)^2\over(1-z_1z_2)^2
(1-z_2z_3)^2(1-z_3z_1)^2} {z_2W(z_2)\over
                               z_1z_3W(z_1^{-1})W(z_3^{-1})}.
\nonumber\eea

   \end{document}